\documentclass[conference]{IEEEtran}
\IEEEoverridecommandlockouts
\pdfoutput=1

\usepackage{cite}
\usepackage{amsmath,amssymb,amsfonts}
\usepackage{graphicx}
\usepackage{textcomp}
\usepackage{xcolor}
\usepackage{orcidlink}
\usepackage[absolute,overlay]{textpos}
\usepackage{booktabs}
\usepackage[linesnumbered,ruled,vlined]{algorithm2e}

\def\BibTeX{{\rm B\kern-.05em{\sc i\kern-.025em b}\kern-.08em
    T\kern-.1667em\lower.7ex\hbox{E}\kern-.125emX}}

\begin{document}

\title{ARRC: Explainable, Workflow-Integrated Recommender for Sustainable Resource Optimization Across the Edge–Cloud Continuum}

\author{
\IEEEauthorblockN{Brian-Frederik Jahnke\orcidlink{0000-0002-7995-6472}}
\IEEEauthorblockA{Fraunhofer ISST\\
Dortmund, Germany\\
brian-frederik.jahnke@isst.fraunhofer.de}
\and
\IEEEauthorblockN{René Brinkhege\orcidlink{0009-0006-8055-0580}}
\IEEEauthorblockA{Fraunhofer ISST\\
Dortmund, Germany\\
rene.brinkhege@isst.fraunhofer.de}
\and
\IEEEauthorblockN{Jan Peter Meyer\orcidlink{0009-0007-4930-5285}}
\IEEEauthorblockA{Fraunhofer ISST\\
Dortmund, Germany\\
jan.peter.meyer@isst.fraunhofer.de}
\and
\IEEEauthorblockN{Daniel Tebernum\orcidlink{0000-0002-4772-9099}}
\IEEEauthorblockA{Fraunhofer ISST\\
Dortmund, Germany\\
daniel.tebernum@isst.fraunhofer.de}
\and
\IEEEauthorblockN{Falk Howar\orcidlink{0000-0002-9524-4459}}
\IEEEauthorblockA{TU Dortmund\\
Dortmund, Germany\\
falk.howar@tu-dortmund.de}
}

\maketitle

\begin{abstract}
Achieving sustainable, explainable, and maintainable automation for resource optimization is a core challenge across the edge–cloud continuum. Persistent overprovisioning and operational complexity often stem from heterogeneous platforms and layered abstractions, while systems lacking explainability and maintainability become fragile, impede safe recovery, and accumulate technical debt. Existing solutions are frequently reactive, limited to single abstraction layers, or require intrusive platform changes, leaving efficiency and maintainability gains unrealized.

This paper addresses safe, transparent, and low-effort resource optimization in dynamic, multi-tenant edge–cloud systems, without disrupting operator workflows or increasing technical debt. We introduce ARRC, a recommender system rooted in software engineering design principles, which delivers explainable, cross-layer resource recommendations directly into operator workflows (such as tickets and GitOps pull requests). ARRC encapsulates optimization logic in specialized, auditable agents coordinated via a shared interface, supporting maintainability and extensibility through transparency and the ability to inspect both recommendations and their rationale.

Empirical evaluation in a multi-region industrial deployment shows that ARRC reduces operator workload by over 50\%, improves compute utilization by up to $7.7\times$, and maintains error rates below 5\%, with most benefits achieved through incremental, operator-approved changes. This demonstrates that explainable, recommendation-based architectures can achieve sustainable efficiency and maintainability improvements at production scale.

ARRC provides an empirically evaluated framework for integrating explainable, workflow-driven automation into resource management, intended to advance best practices for robust, maintainable, and transparent edge–cloud continuum platforms.
\end{abstract} %
\section{Introduction}\label{sec.intro}

Persistent resource waste and operational complexity are major challenges across the modern edge--cloud continuum~\cite{ccit}, affecting domains from IoT and smart cities to mission-critical workloads~\cite{armbrust,shi2016edge}. Chronic overprovisioning and inefficiency are driven by heterogeneous platforms, layered abstractions (e.g., virtualization, containers), and the absence of scalable, explainable, and maintainable resource management~\cite{mell2011nist,satyanarayanan2017emergence,gan2019benchmark,zhang}. 

While autoscalers and predictive optimizers have improved aspects of cloud resource management, most existing solutions remain reactive, operate at a single abstraction layer, or require intrusive changes to production platforms~\cite{satyanarayanan2019edge,casalicchio2019survey,barroso,fu2024alps,lia2024jiagu}. Manual tuning is infeasible at scale, and prior tools either demand costly operator intervention or eliminate human oversight, creating barriers to reliability, maintainability, and adoption~\cite{galantino,autopilot,bsi}.

From a software engineering perspective, these operational inefficiencies are tightly linked to maintainability challenges~\cite{beyer2016sre,adkins2020secure_reliable_software}. As emphasized in SRE and reliability engineering, systems that are not explainable and maintainable become increasingly fragile as they evolve. If operators cannot inspect and understand the rationale behind resource decisions, diagnosing failures and ensuring safe recovery becomes error-prone and slow, especially in complex, distributed architectures. Lack of modularity further complicates adapting systems to new requirements or extending optimization logic to domains such as security and compliance, increasing technical debt \cite{winters2020software}.

This work addresses the problem: \emph{How can we enable safe, transparent, and low-effort resource optimization in dynamic, multi-tenant edge--cloud continuum systems, without disrupting established operator workflows or increasing technical debt?} Our goal is to deliver proactive, cross-layer optimization that integrates seamlessly with production platforms and operator processes, advancing maintainability through transparency and inspectability.

Meeting this challenge requires: (i) coordinating explainable recommendations across VMs, containers, and compliance domains; (ii) integrating with existing operator workflows (e.g., tickets, GitOps pull requests) for safe, auditable, incremental adoption; and (iii) supporting large, dynamic, multi-tenant deployments at scale.

We present ARRC (Automatic Recommender for Resource Configurations), a recommender system grounded in software engineering design principles. ARRC integrates explainable, cross-layer resource recommendations directly into operator workflows, without requiring changes to core infrastructure or user code. Optimization logic is encapsulated in specialized, auditable agents coordinated via a shared interface, supporting maintainability and extensibility through transparency and the ability to inspect and understand recommendations and their rationale.

Evaluated over three months in a multi-region industrial deployment, ARRC achieves up to $7.7\times$ improvement in compute utilization, with most benefit realized from applying only $15\%$ of recommendations. Operator workload for optimization is reduced by $56\%$, while maintaining error rates below $5\%$. In fully autonomous mode, ARRC outperforms Kubernetes’ Vertical Pod Autoscaler (VPA)~\cite{vpa}, reducing vCore usage by over $3\times$ with no observed failures.

\textbf{Contributions:}
\begin{itemize}
    \item An empirically validated architecture for cross-layer, explainable, and maintainable resource optimization in heterogeneous edge--cloud continuum environments;
    \item Design principles for incremental, auditable workflow integration with production platforms;
    \item An empirical case study demonstrating substantial efficiency gains and operator effort reduction at industry scale.
\end{itemize} %
\section{Background and Related Work}
\label{sec:background-related}

\subsection{Key Technical Background}

Resource optimization across the edge--cloud continuum is fundamentally constrained by persistent inefficiencies in resource allocation and utilization. Modern platforms comprise heterogeneous, multi-tenant environments that include virtual machines (VMs), containers, and specialized workloads orchestrated by systems such as Kubernetes and OpenStack~\cite{shi2016edge,kubernetes,openstack,satyanarayanan2017emergence}. Operators are required to specify static resource requests~\cite{tirmazi2020borg,kubernetes,yarn,mesos}, yet typically lack visibility into workload internals due to privacy, compliance, and multi-tenancy boundaries~\cite{mell2011nist,satyanarayanan2017emergence,gan2019,zhang}. Empirical studies indicate that chronic overprovisioning is prevalent, with most VMs and containers sustaining average CPU utilization below 20\%~\cite{delimitrou2014,bashir2021,lang2016,gan2019,lu2014alibaba,reiss2011googletraces,reiss2012googletraces,cortez,zhang}. These patterns are compounded by the predominance of long-lived workloads with stationary resource demand and project-level homogeneity~\cite{cortez,gan2019}.

Technical and organizational challenges are further exacerbated by the lack of explainable, workflow-integrated automation. In practice, operators are often forced to review and enact resource changes manually, increasing the risk of configuration drift, undocumented changes, and technical debt~\cite{beyer2016sre,adkins2020secure_reliable_software}. The absence of transparent rationales for recommendations undermines trust, impedes efficient incident response, and complicates compliance with organizational and regulatory standards.

Workflow integration and explainability have thus emerged as critical requirements for sustainable automation. Explainable recommendations allow operators to inspect, audit, and understand the rationale and projected impact of proposed changes, supporting safe adoption and effective debugging. Seamless integration with established operator workflows, such as tickets and GitOps pull requests, enables incremental, auditable adoption and ensures that automation aligns with organizational processes and compliance expectations.

\subsection{Summary of Related Work}

Early resource management systems, including Borg, Omega, YARN, Mesos, and Kubernetes, introduced scalable scheduling and explicit resource requests~\cite{tirmazi2020borg,schwarzkopf2013omega,yarn,mesos,kubernetes}. These foundational approaches, while robust, are limited by static allocation, underutilization, and the absence of workflow-integrated or explainable automation~\cite{delimitrou2014,gan2019,lang2016,bashir2021}.

Modern techniques have increasingly employed predictive analytics and machine learning to improve utilization, exemplified by systems such as Quasar~\cite{delimitrou2014}, Paragon~\cite{delimitrou2013}, Tarcil~\cite{delimitrou2015}, Sinan~\cite{zhang2}, Seagull~\cite{poppe2020}, and Starburst~\cite{lua2024starburst}. However, these solutions often require intrusive integration, new abstractions, or operate only at a single abstraction layer. Critically, they seldom deliver operator-facing, explainable recommendations or integrate with established workflows, limiting their maintainability and adoption in heterogeneous environments.

Frameworks such as Google Autopilot~\cite{autopilot} offer extensibility but frequently lack cross-layer analytics or auditable, workflow-integrated feedback. Recent advances in software engineering have highlighted the importance of workflow integration and explainability in automation, as seen in AutoRestTest~\cite{AutoRestTest}, RepairAgent~\cite{RepairAgent}, REDII~\cite{REDII}, and AssetHarvester~\cite{AssetHarvester}. These systems demonstrate the value of surfacing actionable feedback within operator workflows and providing transparent, auditable rationales for automated actions, but primarily target domains such as testing, repair, or security, rather than resource optimization.

Persistent gaps remain: most current tools do not unify workflow integration and explainability for resource optimization. The lack of actionable, incrementally adoptable, and auditable recommendations in heterogeneous, multi-tenant environments continues to impede sustainable efficiency and maintainability.

\subsection{Positioning and Contribution}

ARRC directly addresses these limitations by delivering a workflow-integrated, explainable resource optimization system for heterogeneous edge--cloud environments. The architecture is grounded in established software engineering and reliability principles, with the following distinguishing features:
\begin{itemize}
    \item \textbf{Operator workflow integration}: Recommendations are surfaced directly through tickets and pull requests, supporting incremental, auditable, and traceable adoption.
    \item \textbf{Explainability}: Each recommendation provides transparent rationale, supporting evidence, and projected impact, enabling operators to review, audit, and approve changes safely.
    \item \textbf{Cross-layer analytics}: ARRC coordinates recommendations across VMs, containers, and compliance/security domains without requiring disruptive integration or new abstractions.
    \item \textbf{Empirical validation}: Evaluation in a multi-region industrial deployment demonstrates up to $7.7\times$ utilization improvement and over 50\% operator workload reduction, with low error rates and most benefits achieved by implementing a prioritized subset of recommendations.
\end{itemize}
The workflow-centric, explainable approach underlying ARRC is broadly applicable to other domains requiring operator-centric, sustainable automation, including cost, security, and reliability optimization.

\section{Approach: Automatic Recommender for Resource Configuration}
\label{sec:approach}

This section describes the design and implementation of ARRC.

\subsection{High-Level Architecture}
\label{sec:architecture}

\begin{figure}[t]
    \centering
    \includegraphics[width=0.92\linewidth]{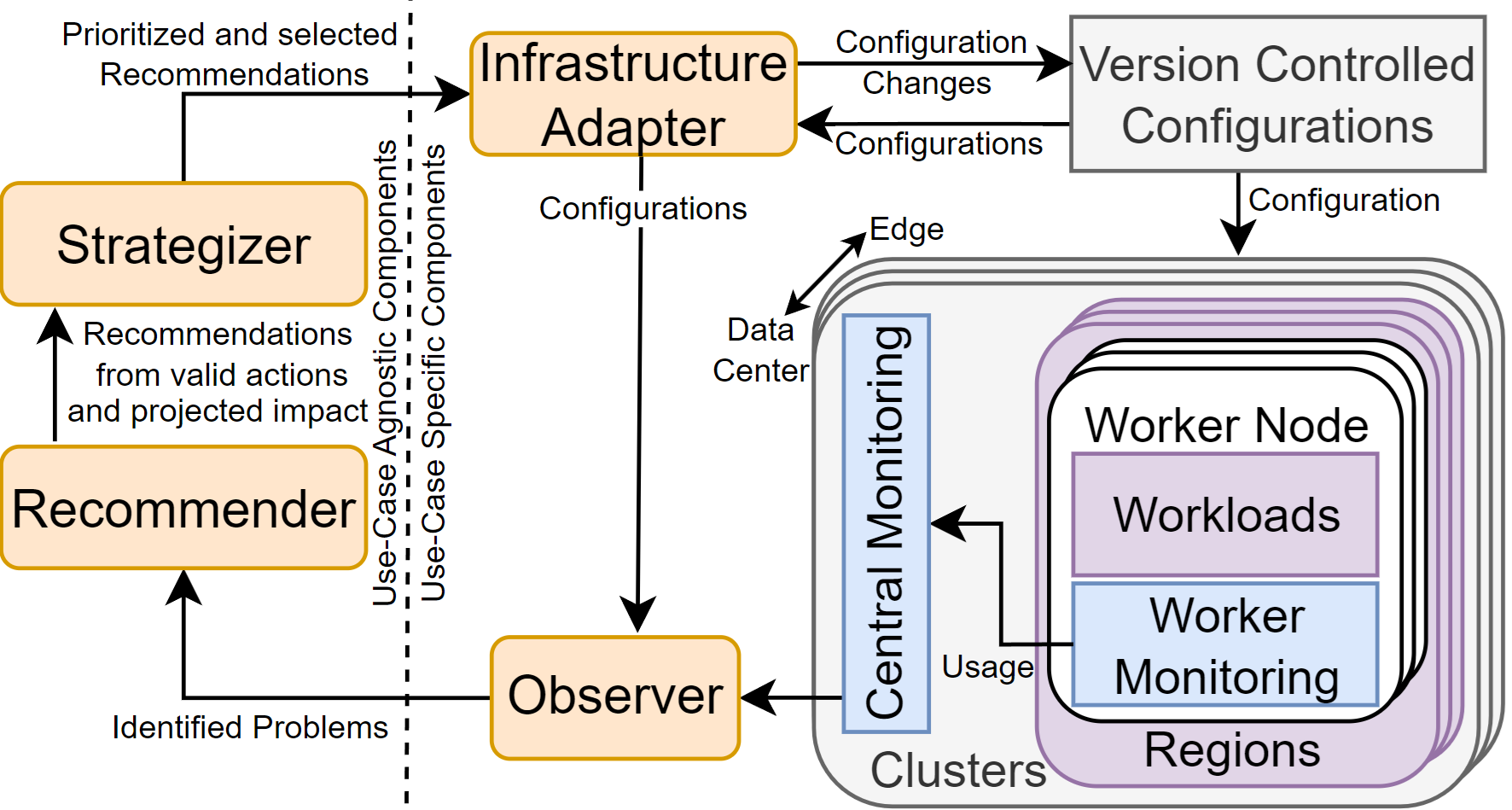}
    \caption{ARRC data flow: The Observer collects and normalizes telemetry, posting versioned observations to the blackboard. Agents analyze state and propose recommendations; the Strategizer sequences actions; recommendations are surfaced to operator workflows via the Infrastructure Adapter.}
    \label{fig:arrc-dataflow}
\end{figure}

Figure~\ref{fig:arrc-dataflow} provides an overview of ARRC, which delivers explainable, workflow-integrated resource optimization across heterogeneous edge--cloud environments. The system follows a modular, agent-based design (see Table~\ref{tab:arrc-components}) coordinated via a versioned blackboard. The workflow comprises four phases: (1) ingestion and normalization of resource state and telemetry from infrastructure adapters for Kubernetes, OpenStack, Prometheus, and file-based sources; (2) analysis and recommendation generation by specialized agents employing time-series forecasting and heuristic/statistical analysis; (3) prioritization, conflict resolution, and sequencing of recommendations using a goal-oriented Strategizer; and (4) surfacing of top-ranked recommendations into operator workflows (merge requests, tickets, issues), with a configurable cap (default: 10 per week). The architecture supports both centralized and distributed deployments, using atomic transactions and versioning to ensure consistency and auditability, and is designed for operator-in-the-loop adoption.

\begin{table}[h]
\centering
\caption{ARRC Core Components and Roles}
\begin{tabular}{lp{5.5cm}}
\toprule
Component & Role \\
\midrule
Observer & Ingests and normalizes telemetry, versions \texttt{Observation} records for all resources. \\
Recommender Agents & Analyze observations and generate explainable, evidence-backed recommendations (e.g., RightSizing, Security). \\
Strategizer & Prioritizes, sequences, and coordinates actions; resolves agent conflicts using GOAP-based prioritization and dynamic weighting. \\
Workflow Adapter & Surfaces recommendations to operator workflows (merge/pull requests, tickets, issues); manages branch/commit lifecycle and logs operator feedback. \\
Blackboard & Versioned, atomic data store for observations and recommendations; enables traceability, auditability, distributed/global deployments. \\
\bottomrule
\end{tabular}
\label{tab:arrc-components}
\end{table}

\subsection{Core Components and Subsystems}
\label{sec:components}

\subsubsection{State Collection and Data Model}
The Observer subsystem collects resource state and telemetry using adapters for Kubernetes, OpenStack, Prometheus, and file-based sources. Each adapter maps platform-specific resource data into standardized, versioned \texttt{Observation} records, which include resource descriptors, timestamps, and relevant metrics or scan results. Observations are atomically written to the blackboard, a versioned key-value store (e.g., etcd~\cite{etcd}), supporting both global and per-cluster deployments with full auditability.

\subsubsection{RightSizing Recommender Agent}
\label{sec:rightsizing}
The RightSizing agent in ARRC is responsible for generating explainable, evidence-backed recommendations to optimize resource allocations for virtual machines and containers. Its main function is to analyze both historical and forecasted utilization and to propose configuration changes that bring provisioned resources into alignment with actual and predicted demand.

The analysis begins by grouping versioned observation records by resource handle, such as a VM or container identifier. Each observation aggregates time-series telemetry over a rolling, operator-configurable window (default: seven days), capturing CPU and memory requests, limits, and actual usage. Observations are enriched with contextual metadata, including resource flavor, project, owner, container image, and instance identifiers, and they are atomically persisted in the blackboard for traceability.

For each resource, the agent computes key statistical features: the 95th percentile (P95), minimum, maximum, and slack (difference between allocated and used resources) for CPU and memory. Persistent breaches of operator-defined lower or upper utilization thresholds are identified by examining these historical statistics. To improve robustness under both stationary and variable workloads, the agent applies a hybrid forecasting pipeline: Prophet and ARIMA models are executed in parallel, and a heuristic model selector, based on time-series complexity measures such as sample entropy and variance (see~\cite{darts}), determines the most suitable model for each case. The forecast window is operator-configurable, defaulting to seven days.

Resources identified as candidates for rightsizing are matched against the available resource catalog to determine the optimal target configuration. At a minimum, CPU and memory requirements are matched directly. If additional configuration dimensions (e.g., storage, node architecture, resource tags) are present, the agent computes the Euclidean distance across all relevant key-value pairs, enabling multi-attribute optimization. To ensure operational safety, a configurable resource buffer (default: 10\%) is added to the recommended allocation, absorbing measurement noise and short-term spikes. If the newly selected configuration is identical to the current allocation, no recommendation is emitted, avoiding unnecessary churn.

Each recommendation produced by the agent includes a configuration patch (in YAML or JSON), a quantified impact analysis (projecting changes in cost, energy, utilization, and reliability, normalized to cluster capacity), and a detailed rationale. The impact analyzer classifies expected improvements by type, such as reliability, performance, and cost efficiency, and all underlying data, input observations, statistical summaries, and forecast values are referenced explicitly in the recommendation object. Every recommendation is versioned and includes a unique identifier for auditability. Operator feedback is integrated directly: recommendations are surfaced as merge/pull requests or tickets, and their approval, rejection, or modification is logged. If a recommendation is rejected within the current reporting window, it is not reissued until the next window, preventing redundant proposals and reducing notification fatigue. While explicit online learning is not performed, historical operator feedback is retained for future policy analysis and tuning at the agent level.

For example, if a VM with 8 vCPUs and 32 GiB RAM sustains average CPU utilization of 1.4\% and memory utilization of 10\% for several months, the RightSizing agent applies Prophet and ARIMA forecasting, detects persistent underutilization, and uses multi-key nearest-neighbor selection to recommend downsizing to a 4C/16GiB instance. The agent quantifies expected improvements in cost, utilization, and energy, and emits a recommendation referencing all supporting evidence and rationale for operator review.

This agent-centric design addresses the opacity and lack of explainability in prior autoscalers and black-box optimizers. By combining robust forecasting, catalog-aware optimization, and systematic feedback integration, the RightSizing agent ensures recommendations are actionable and auditable. The modular and explainable interface also supports rapid extension to new resource types and requirements, facilitating maintainable, operator-centric resource optimization at scale.

\subsubsection{Security Recommender Agent}

The Security agent in ARRC automates the detection and remediation of Kubernetes security compliance violations, providing explainable and actionable recommendations that integrate seamlessly with operator workflows. The agent is realized via the \texttt{KubescapeObserver} and \texttt{KubescapeRecommender} components, collaborating to process the results of periodic cluster-wide security scans.

The workflow begins with the \texttt{KubescapeObserver}, which schedules and executes security scans using Kubescape~\cite{kubescape} for security frameworks such as NSA-CISA (U.S. government Kubernetes security guidance)~\cite{NSA_CISA_Benchmark} and CIS Benchmarks (industry-standard security configuration guidelines)~\cite{cisBenchmark}. Scan results, typically produced as HTML reports, are parsed to extract failed compliance checks and capture structured attributes such as check name, resource name and kind, namespace, severity, documentation links, and remediation hints. Each failed check is encoded as a uniquely named, timestamped, and versioned \texttt{Observation} record in the blackboard, enabling traceability and downstream processing.

Upon creation of new security observations, the \texttt{KubescapeRecommender} agent processes each failed check by synthesizing a contextualized recommendation name and constructing a detailed prompt for a large language model (LLM), such as Llama 3~\cite{grattafiori2024llama3herdmodels}. The prompt is built by serializing the observation as JSON and scraping the linked documentation to extract the most relevant explanatory text (e.g., via the ".rm-Article" CSS selector). The resulting prompt typically follows this structure:

\begin{quote}
I did run a kubescape scan and it identified the following problem. Please give me a step by step guide on how to resolve this:

\{\{.kubescanreport\}\}

\{\{.related\_web\_doc\}\}.
\end{quote}

The LLM is invoked with a strict timeout to ensure responsiveness. The returned remediation steps are recorded as the explanation in a new \texttt{Recommendation} object, which also includes severity, rationale, documentation links, and a reference to the originating observation. Each recommendation is fully versioned, with all supporting evidence, input scan, extracted documentation, and LLM-generated output, directly linked for auditability. This ensures that every recommendation is explainable and empirically grounded, allowing operators and auditors to reconstruct both the evidence and reasoning behind each proposed remediation.

Operator feedback is integrated at the agent level: recommendations are surfaced as merge/pull requests or tickets, and their approval, modification, or rejection is logged. Recommendations rejected during the current reporting window are not reissued until the next window, reducing redundancy and notification fatigue. While the agent does not currently perform explicit online learning, the accumulated feedback and logs enable iterative improvement of prompt construction, documentation extraction, and guidance quality.

For example, if a Kubescape scan identifies a pod lacking a required security context, the agent parses the failed check, scrapes the relevant documentation, and queries the LLM for a remediation guide. The resulting recommendation provides a prioritized rationale, documentation links, and actionable steps for the operator, with all information referenced to the original scan and documentation. If approved, the remediation is enacted and logged for compliance; if rejected, the recommendation is suppressed until the next reporting window.

The Security agent’s design directly addresses the lack of explainability and automation in prior compliance tooling by surfacing transparent, auditable, and operator-facing remediation steps. Its modular, evidence-driven approach facilitates adaptation to evolving security benchmarks and supports maintainable, operator-centric management of compliance in production settings.

\subsubsection{Strategizer: Prioritization, Conflict Resolution, and Goal Reasoning}

\begin{figure}[t]
    \centering
    \includegraphics[width=0.92\linewidth]{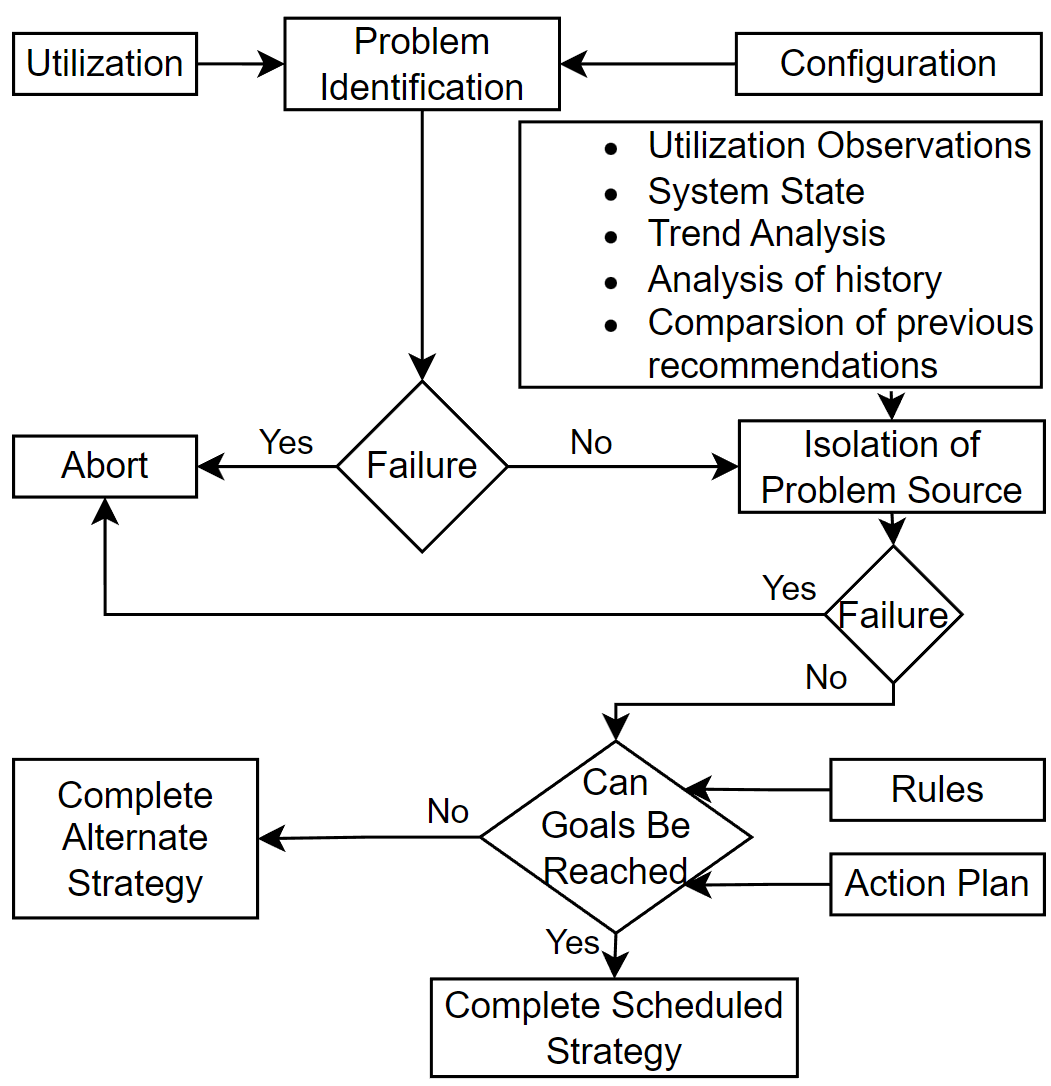}
    \caption{Cluster optimization logic: The Strategizer orchestrates actions from multiple agents, evaluates their combined impact, and sequences changes for safe, incremental adoption.}
    \label{fig:arrc-controlflow}
\end{figure}

The Strategizer is responsible for globally prioritizing, sequencing, and filtering all recommendations generated by ARRC’s agents. As depicted in Figure~\ref{fig:arrc-controlflow}, the Strategizer operates after agents have independently analyzed the blackboard and emitted their recommendations. Its primary role is to ensure that only the most impactful and context-appropriate recommendations are surfaced to operator workflows, while maintaining consistency, fairness, and alignment with operator-defined objectives.

The Strategizer maintains an internal queue of candidate recommendations, each annotated with metadata including origin agent, affected resource, estimated impact (e.g., projected improvements in reliability, cost, utilization, or compliance), and operator feedback history. To rank and select recommendations, the Strategizer applies a hybrid static/dynamic weighting system based on operator-configured priorities for criteria for software qualities according to ISO25010~\cite{iso_25010}: reliability, performance, security, cost, and sustainability. Static priorities define the baseline order (e.g., reliability $>$ performance $>$ security $>$ cost $>$ sustainability), while dynamic weights are computed at runtime based on the current distance to operator-specified targets for each objective. This enables the system to emphasize actions that most directly close gaps in key metrics.

For conflict resolution, the Strategizer implements goal-oriented action planning (GOAP)~\cite{orkin2006gdc} to globally resolve situations where multiple agents propose changes to the same resource or to interdependent resources. It identifies and clusters conflicting or overlapping recommendations and applies an operator-specified resolution policy (e.g., reliability overrides cost, or performance trumps sustainability). When agent recommendations are mutually exclusive or could cause cascading effects, the Strategizer selects the action that best aligns with current global priorities, suppressing or deferring others. This ensures safe, incremental adoption and prevents unintended side effects.

To manage operator and developer workflow load, the Strategizer enforces a configurable cap on surfaced recommendations (default: top 10 per reporting window, operator-adjustable). Recommendations are sorted by combined impact and priority; only the highest ranked are surfaced as actionable merge requests, tickets, or issues. All recommendations, including those not surfaced, remain in the blackboard for auditability, traceability, and future analysis.

The Strategizer does not generate new recommendations or alter agent-level analysis logic. Its design is intentionally decoupled from agent internals to facilitate modularity and maintainability. By centralizing prioritization, conflict resolution, and sequencing, the Strategizer enables ARRC to achieve multi-objective, globally consistent optimization, while supporting transparent, operator-in-the-loop decision making. This architecture addresses prior limitations by ensuring that automation is safe, explainable, and aligned with business and reliability goals across complex, multi-agent resource management scenarios.

\subsubsection{Workflow Integration and Operator Feedback}
\begin{figure}
    \centering
    \includegraphics[width=1.0\linewidth]{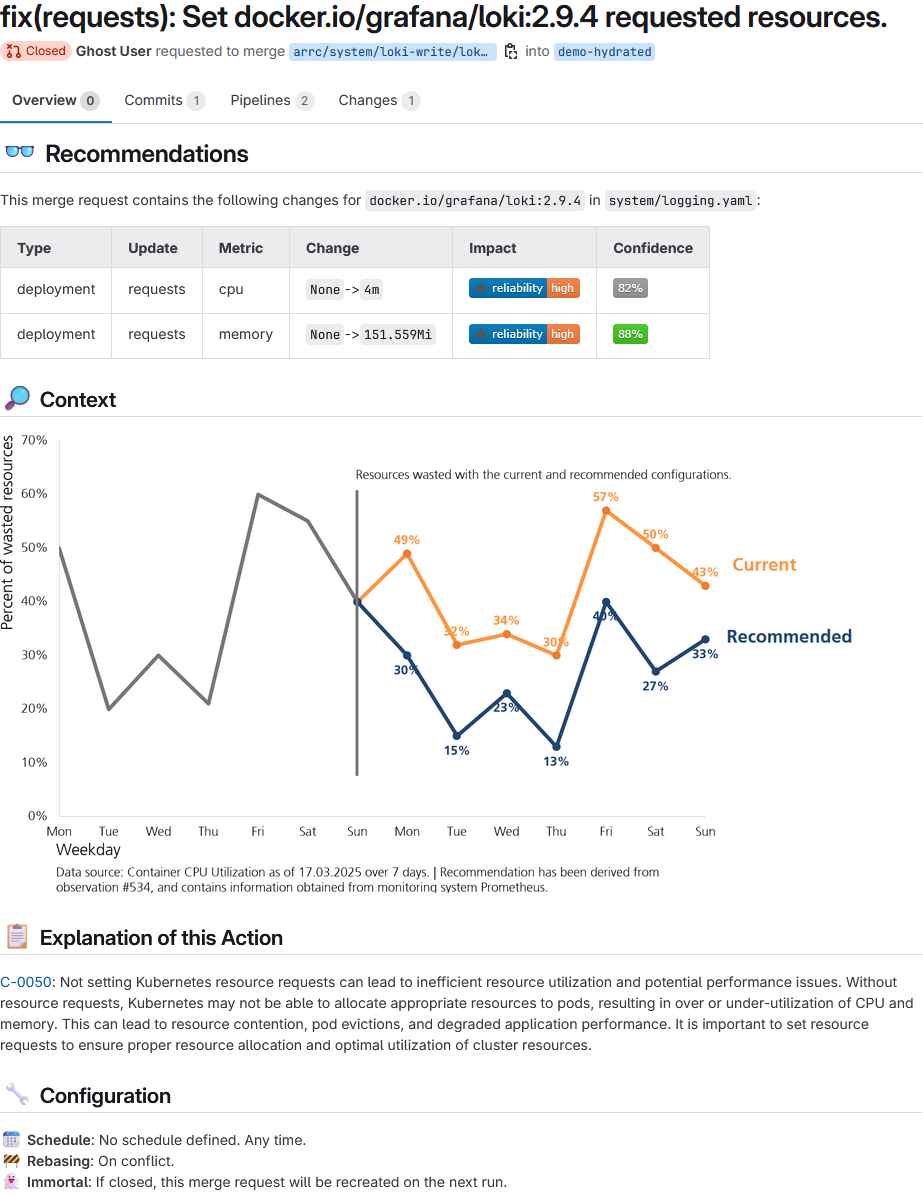}
    \caption{An example merge request for changing the requested resources of a container. The commit contains the described changes directly in the repository.}
    \label{fig:merge-request}
\end{figure}

Figure~\ref{fig:merge-request} illustrates an example of how ARRC surfaces a resource optimization recommendation directly into operator workflows: a merge request is generated to change the requested resources of a container, with the commit containing the described changes in the repository. This concrete example highlights the system’s focus on transparent, traceable, and incremental adoption of optimization actions through familiar developer tools.

The Infrastructure Adapter is responsible for translating each Recommendation, as prioritized by the Strategizer, into the concrete output format and system required by the target operator workflow. This entails mapping the internal, platform-agnostic recommendation objects into specific formats (e.g., GitHub or GitLab merge requests, pull requests, issues, or ticketing system entries) and generating the corresponding configuration changes (such as YAML/JSON patches or compliance scripts). By abstracting the translation process, the Infrastructure Adapter enables seamless integration with diverse operational platforms and ensures that each surfaced recommendation is both actionable and auditable within established workflows. Each output includes a machine- and human-readable summary, rationale, supporting evidence, and direct links to best-practice documentation, maintaining transparency and traceability across the entire optimization lifecycle.

Operator actions, including approval, rejection, or modification, are logged and versioned. Recommendations that are rejected within a reporting window are not resurfaced until the next window, reducing redundancy and notification fatigue. All operator feedback is retained for future analysis and policy tuning. The workflow integration supports deployment at per-cluster, per-region, or global levels, with ARRC optimizing all accessible resources transparently and consistently across distributed environments.

This design enables operator-in-the-loop optimization, ensures traceability and compliance, and provides the empirical foundation for ARRC’s evaluation of operator effort reduction and adoption in practice (see Section~\ref{sec:evaluation}).

\subsubsection{Implementation and Deployment}
ARRC is implemented in Go with modular adapters and uses a versioned key-value blackboard for atomic, consistent operations. Data serialization employs JSON and YAML; all workflow and API interactions leverage standard REST APIs. Automated tests cover all major modules. Deployment topology is operator-controlled, with ARRC optimizing all accessible clusters and regions.

\subsection{Example Scenario}
\label{sec:scenario}

Consider a VM with 8 vCPUs and 32~GiB RAM sustaining average CPU utilization of 1.4\% and memory utilization of 10\% over several months. The Observer collects and normalizes this telemetry, posting an \texttt{Observation} to the blackboard. The RightSizing agent analyzes usage and forecasts demand, detects persistent underutilization, and selects a 4C/16GiB instance via nearest-neighbor search. The agent quantifies projected impact and emits a Recommendation. Table~\ref{tab:arrc-sample-recs} provides example recommendations generated by both agents which can be surfaced to the user as merge request (e.g. see Figure~\ref{fig:merge-request}).

\begin{table}[ht]
\caption{Example Recommendations by Instantiated Agents}
\label{tab:arrc-sample-recs}
\centering
\begin{tabular}{p{1.1cm}p{2.2cm}p{1.8cm}p{2.1cm}}
\toprule
Agent & Input (Observation) & Recommendation & Projected Impact \\
\midrule
RightSizing &
VM avg CPU=1.4, m1.large &
Downsize to m1.medium (4C/16GiB) &
Lower cost, higher util., lower energy use \\
Security &
Policy scan: min 8GiB RAM required &
Enforce min memory 8GiB; set K8s resource limits for compliance &
Compliance, workload isolation, possible overprovision \\
\bottomrule
\end{tabular}
\end{table}

\subsection{Design Rationale}
\label{sec:rationale}

ARRC’s architecture targets limitations in explainability, workflow integration, scalability, and operational overhead found in prior resource management solutions. The agent-based structure supports extensibility and maintainability, enabling new optimization and security domains without disruptive rewrites. The versioned blackboard ensures atomicity, auditability, and consistency for all state changes, suitable for both centralized and distributed deployments. Explainability is integral: recommendations explicitly include input data, rationale, and supporting evidence, and are surfaced directly to operator workflows for transparent, auditable review, addressing a key shortcoming of autoscalers and black-box ML optimizers. The forecasting pipeline (Prophet, ARIMA, heuristics) and nearest-neighbor configuration search provide operational robustness and flexibility. Goal-oriented action planning in the Strategizer enables multi-objective prioritization and conflict resolution, balancing reliability, performance, security, cost, and sustainability in line with operator policy. The configurable cap on surfaced recommendations prevents overload and focuses attention on high-impact changes. ARRC favors operator-in-the-loop over full autonomy to ensure safety and transparency, and integrates with existing platforms using standard APIs. Versioned, atomic updates and comprehensive logging guarantee consistent, auditable operation in distributed settings, supporting both incremental and global optimization, compliance, and safe rollback.
\section{Evaluation}
\label{sec:evaluation}

This section evaluates ARRC’s effectiveness in improving resource utilization, reducing operator effort, and maintaining reliability in a production, multi-region edge--cloud continuum deployment. All results are derived from the system implemented as described in Section~\ref{sec:approach}. We address the following research questions:
\\
\textbf{RQ1:} How effective is ARRC at improving resource utilization and reducing resource waste? \\
\textbf{RQ2:} What is the impact of ARRC on operator effort and workflow? \\
\textbf{RQ3:} How does ARRC compare to standard baselines, such as Kubernetes VPA, in fully autonomous mode? \\
\textbf{RQ4:} What is the marginal impact of partial or ablated adoption of ARRC recommendations?

\subsection{Evaluation Environment}

ARRC was evaluated in a real-world production environment that combines OpenStack and Kubernetes to support a diverse range of tenants and workloads, including R\&D, IT, and external clients, across multiple geographic regions. These deployments operate under practical constraints such as intermittent connectivity and strict latency requirements. All workloads are treated as opaque third-party applications, and operators have only limited visibility into their internals, reflecting typical privacy and multi-tenancy boundaries found in industry settings.

\begin{figure}[t]
    \centering
    \includegraphics[width=\linewidth]{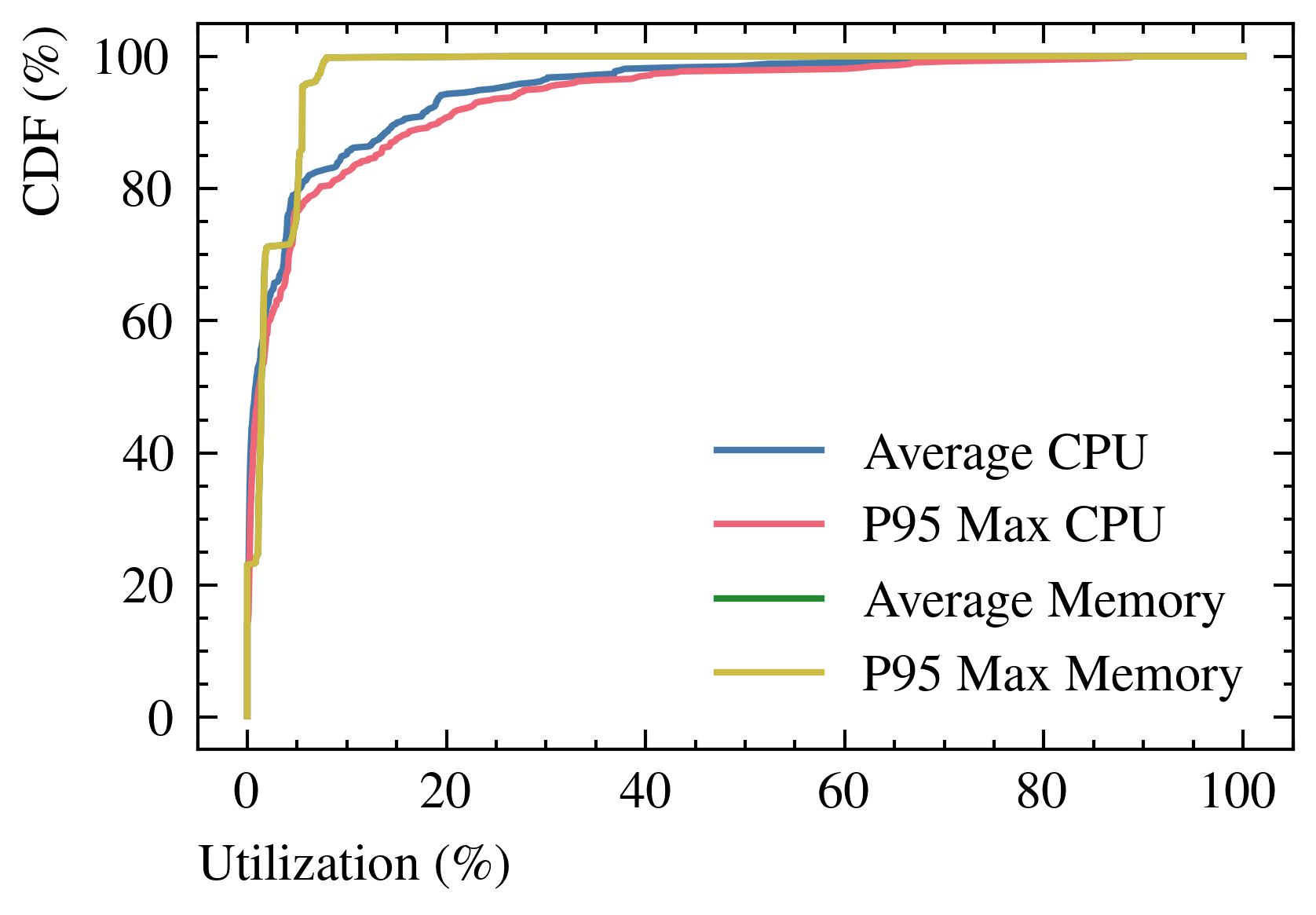}
    \caption{\textbf{CPU and Memory Utilization.} Most VMs are significantly underutilized, both on average and at peak.}
    \label{fig:cpu_cdf}
\end{figure}

\begin{figure}[t]
    \centering
    \includegraphics[width=1.0\linewidth]{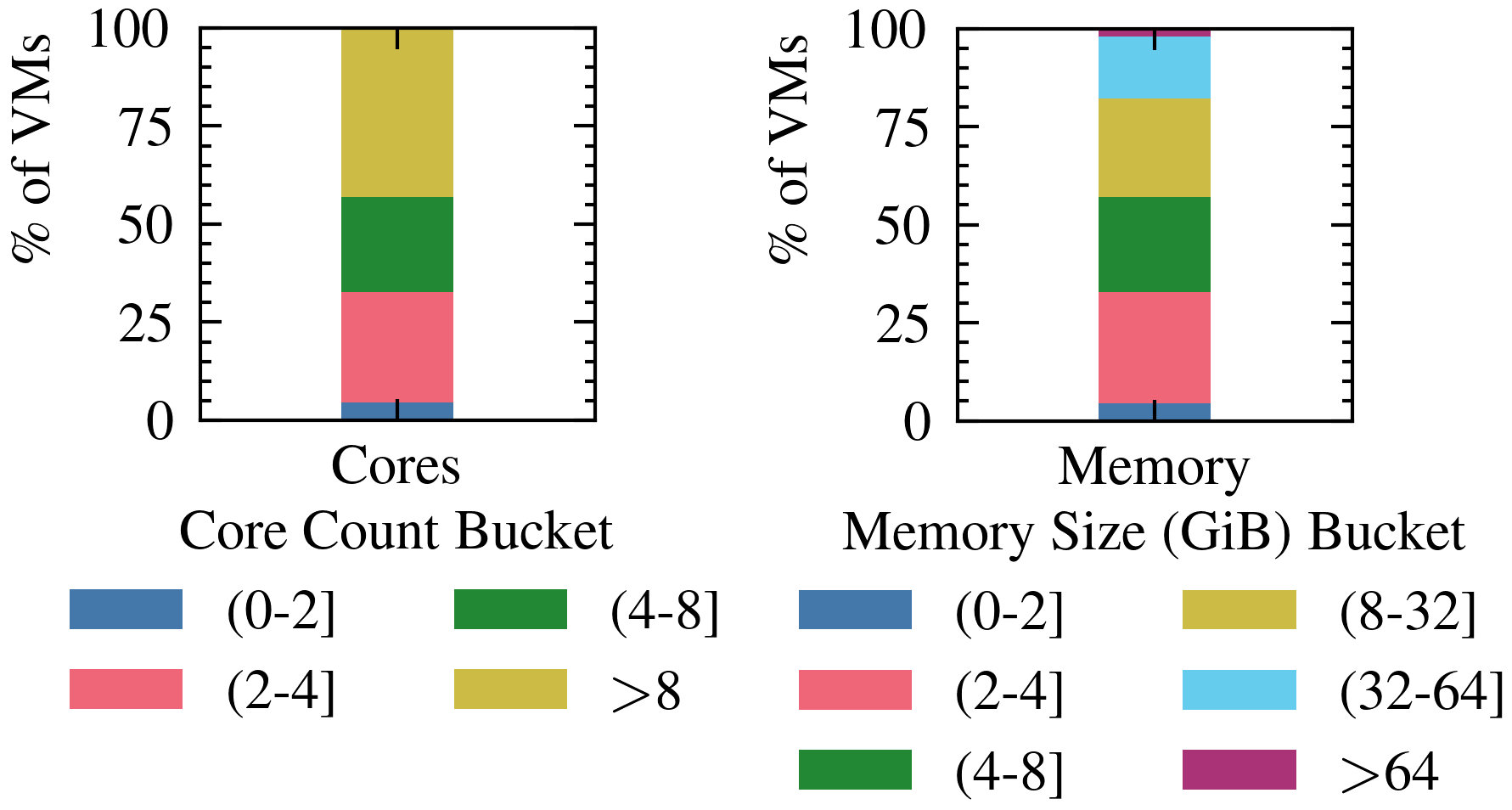}
    \caption{\textbf{VM Core and Memory Distribution.} The majority of VMs have eight or fewer CPU cores and less than 32~GiB memory, allowing flexible placement across edge and cloud sites.}
    \label{fig:vm_cpu_dist}
\end{figure}

To characterize the scale and complexity of the optimization challenge, we analyzed a three-month operational dataset (July--September 2024) covering 528 unique VMs. Each record included identifiers, VM sizing, and fine-grained (five-minute) resource utilization. Following the Resource Central methodology~\cite{cortez}, this analysis revealed several persistent patterns:

\vspace{0.5em}
\begin{itemize}
    \item \textbf{Chronic underutilization and overprovisioning:} Platform-wide, resource consumption grew only modestly, while average CPU utilization for 90\% of VMs was 14.5\%, with 95th-percentile maxima rarely exceeding 19\%. 85\% of VMs operated below 10\% CPU utilization, and 80\% were long-lived. Memory usage showed similar trends.
    \item \textbf{Long-tailed usage:} Just 5\% of VMs accounted for more than half of total CPU hours, resulting in pronounced resource skew.
    \item \textbf{Stationarity and predictability:} Over 95\% of workloads exhibited less than 15\% week-to-week variance in utilization, and nearly all remained stable for at least two weeks (see Figure~\ref{fig:cpu_cdf}).
    \item \textbf{Project-level homogeneity:} Nearly 70\% of projects relied on a single machine type, with low intra-project variance, suggesting that group-level optimization has high potential.
    \item \textbf{Sensible sizing and placement:} Most VMs were provisioned with eight or fewer cores and less than 32~GiB RAM (Figure~\ref{fig:vm_cpu_dist}), a profile well-suited to both edge and cloud sites. Only 2\% exceeded 64~GiB, limiting deployment flexibility.
    \item \textbf{Rare vertical scaling:} Vertical scaling actions were rare; most adaptation occurred through horizontal scaling or replacement. Resource starvation events (CPU throttling, OOM) were both below 1\%.
\end{itemize}
\vspace{0.5em}

These empirical findings highlight persistent overprovisioning, stable usage patterns, and high homogeneity, all of which motivate the need for proactive, explainable, and maintainable optimization. Manual rightsizing is infeasible at this scale, and existing tools do not leverage the observed stability and group-level uniformity.

A typical scenario illustrates the opportunity: a VM with 8 vCPUs and 32~GiB RAM, which averages just 1.4\% CPU and 10\% memory use for months, is rarely reviewed manually. ARRC's RightSizing agent can automatically detect this persistent underutilization, forecast continued low demand, and generate a clear, actionable recommendation to downsize the VM, surfacing this to the operator as an auditable ticket or GitOps pull request. This process, rooted in explainable, modular automation and seamless workflow integration, enables transparent, scalable resource optimization in complex, heterogeneous environments.

Taken together, these workload characteristics directly motivate ARRC’s architectural choices: modular agent specialization for extensibility and maintainability, explainable recommendations for safe adoption, and integration with established workflows for persistent, auditable operational improvement.

\subsection{Recommendation Quality and Optimization Outcomes}

\begin{figure*}[t]
    \centering
    \includegraphics[width=\linewidth]{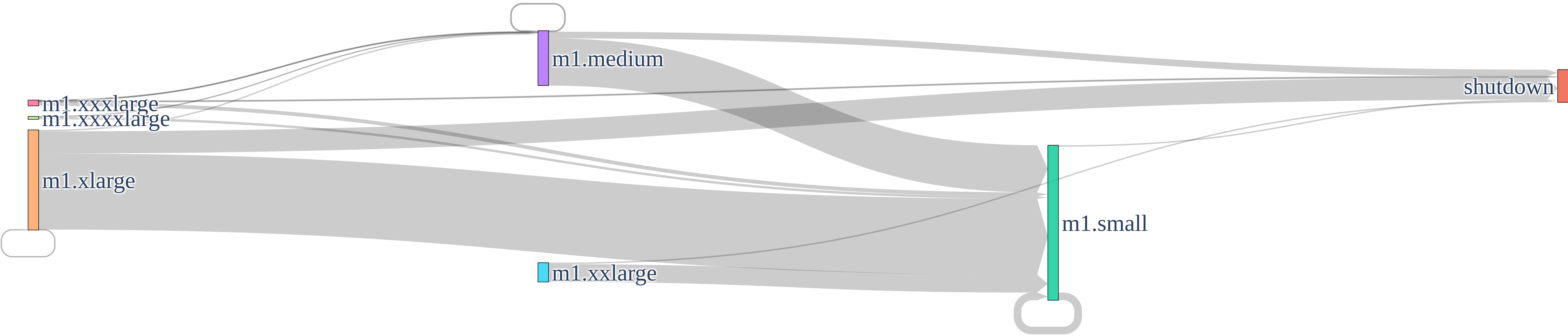}
    \caption{\textbf{Recommended Platform Changes.} Sankey diagram visualizing ARRC's recommended changes from one instance type to another. Most machines are oversized even with 15\% tolerated wastage, so most recommendations reduce cost rather than affect performance or reliability. Arrows from a machine type to itself indicate correctly configured machines (no recommendation generated).}
    \label{fig:recommended_changes}
\end{figure*}

ARRC’s recommendation pipeline prioritizes impactful changes while minimizing operator review effort. Over three months, ARRC analyzed utilization data for 528 VMs, using two months as historical context for forecasting. The system identified that $95.2\%$ of VMs were candidates for downscaling, with only $4.8\%$ correctly sized and $17.1\%$ suspected idle. Figure~\ref{fig:recommended_changes} visualizes these recommendations as a Sankey diagram: most machines are oversized, and most recommendations focus on downsizing, with almost no upscaling required, and notably nearly all machines requiring a change.

\begin{figure}[t]
    \centering
    \includegraphics[width=1.0\linewidth]{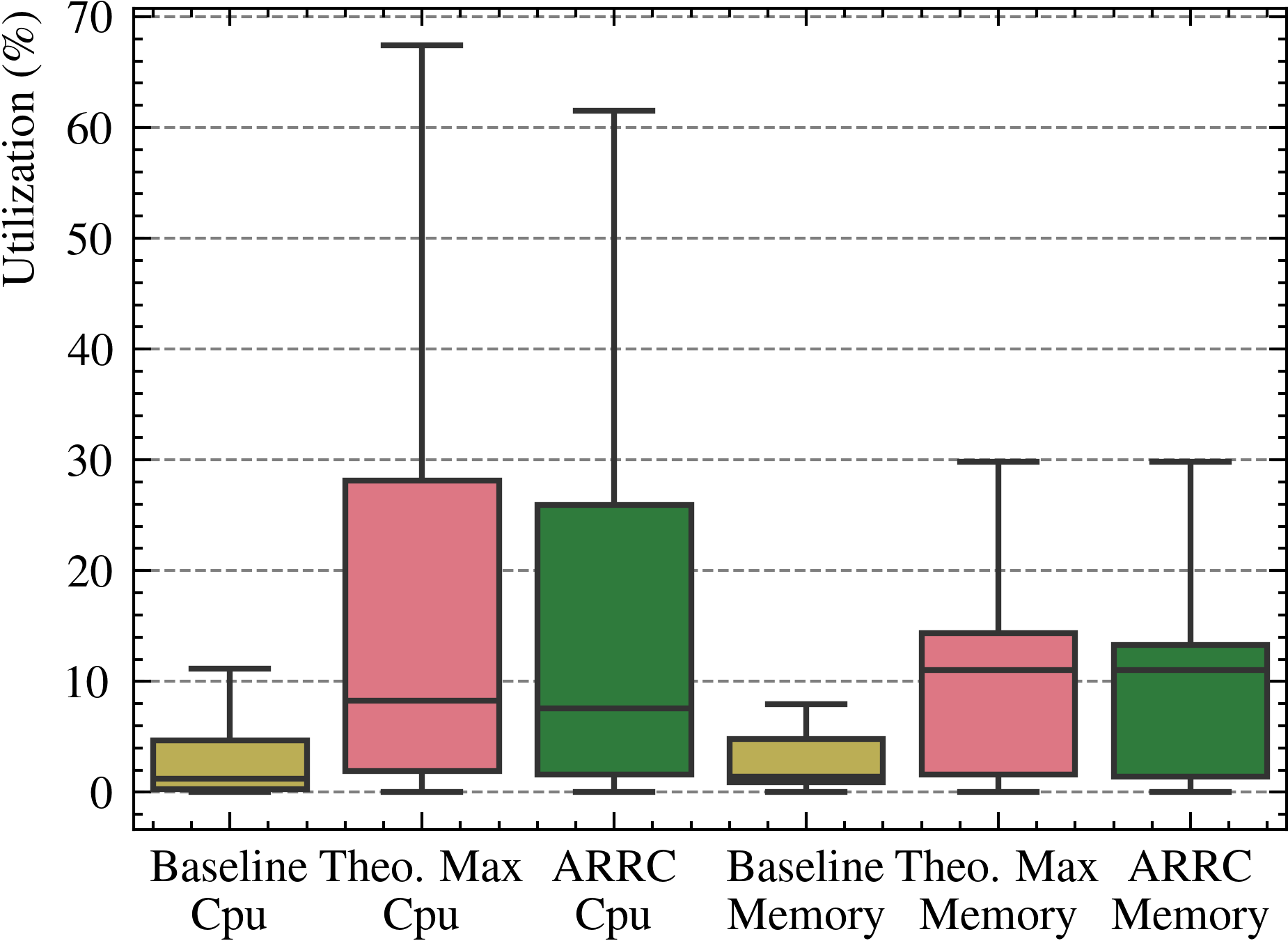}
    \caption{\textbf{Utilization of CPU and memory:} Baseline, theoretical maximum, and ARRC recommendations. ARRC approaches theoretical maximums even during live workload changes.}
    \label{fig:utilization_comparison}
\end{figure}

Applying ARRC’s recommendations led to substantial utilization gains. As shown in Figure~\ref{fig:utilization_comparison}, at the 25th percentile, CPU and memory utilization increased by factors of $6.9$ and $1.6$, respectively, nearly matching the theoretical optima ($7.7$ and $1.6$), where resources are always correctly configured. These results are based on live production metrics from the evaluated deployment (see Section~\ref{sec:approach}).

\begin{figure}[t]
    \centering
    \includegraphics[width=1.0\linewidth]{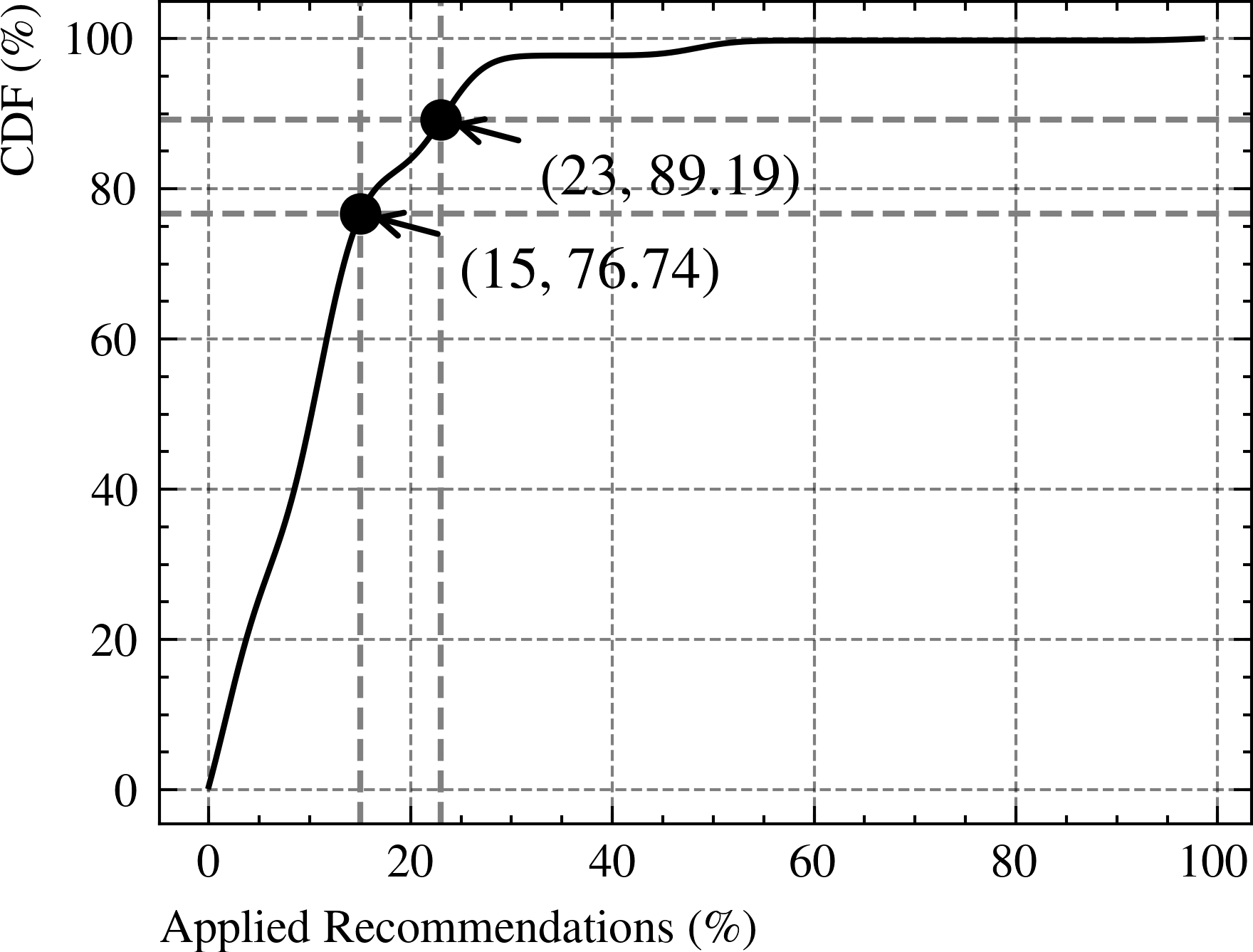}
    \caption{\textbf{Optimization against applied recommendations:} Incremental implementation of prioritized recommendations yields most of the benefit early, demonstrating ARRC's efficient focus.}
    \label{fig:ablation_curve}
\end{figure}

\textbf{Ablation Highlight:} As illustrated in Figure~\ref{fig:ablation_curve}, implementing just the top $15\%$ of prioritized recommendations captured $79\%$ of the attainable optimization benefit. The system converges rapidly, with $91\%$ of total optimization achieved after only $23\%$ of recommendations are applied. This demonstrates that ARRC efficiently focuses operator attention for maximal impact.

\subsection{Operator Effort and Workflow Impact}

Operator workload was measured using ticketing and GitOps logs from the same system. Prior to ARRC, operators manually generated an average of 38 resource optimization tickets per month, spending 25–30 minutes per ticket, for a total of 19 hours monthly. After ARRC deployment, operator-generated tickets decreased by $61\%$, while ARRC itself generated 72 recommendations per month. As reported in Table~\ref{tab:toil}, operators reviewed and approved these in a mean of 7 minutes per ticket, reducing monthly time to $8.4$ hours, a $56\%$ reduction. ARRC’s prioritization enabled operators to focus on high-impact changes first.

\begin{table}[t]
\centering
\small
\caption{Operator Toil Reduction: Before and After ARRC}
\begin{tabular}{@{}lrr@{}}
\toprule
 & Pre-ARRC & With ARRC \\
\midrule
Optimization tickets/month & 38 (manual) & 72 (ARRC-gen) \\
Operator time/ticket (min) & 27 & 7 \\
Total operator time/month (h) & 19 & 8.4 \\
\bottomrule
\end{tabular}
\label{tab:toil}
\end{table}

\subsection{Reliability and Safety}

A recommendation is deemed erroneous if, in the subsequent week, actual peak utilization diverges by more than $15\%$ from the recommended value (see Section~\ref{sec:rightsizing} for step-by-step algorithm description). With three weeks of historical data, ARRC maintained a $3.1\%$ error rate, all single-sided threshold violations, with no observed application failures or service degradations. High-risk recommendations are flagged for operator review; highly dynamic workloads are delegated to reactive autoscalers.

\subsection{Resource and Operational Costs}

ARRC is lightweight, with an application footprint of $42$MB and storage requirements of approximately $16$KB per VM. Network transfer per VM is under $1$KB using cached results, supporting scalability for both large and small clusters.

\subsection{Autonomous Baseline: ARRC vs.\ VPA}

ARRC was compared to Kubernetes’ Vertical Pod Autoscaler (VPA) in a fully autonomous, operator-out-of-the-loop mode, as described in Section~\ref{sec:approach}, where all recommendations are directly and automatically accepted. Both systems applied changes automatically, with ARRC enforcing additional security and compliance constraints. Table~\ref{tab:vcore_usage} shows that ARRC reduced total cluster vCore usage by over $3\times$ compared to VPA, achieving $13$ wasted vCores versus $46$ for VPA and $239$ for static allocation. No application errors or crashes were observed in either case.

\begin{table}[t]
\centering
\caption{Total vCore Usage with No Scaling, VPA, and ARRC (Auto Mode)}
\label{tab:vcore_usage}
\begin{tabular}{lc}
\toprule
Scaling Policy    & Total vCores Wasted \\
\midrule
No Scaling        & 239 \\
VPA (Auto)        & 46 \\
ARRC (Auto)       & 13 \\
\bottomrule
\end{tabular}
\end{table}

\subsection{Summary}

ARRC delivers substantial efficiency gains and operator toil reductions, with low error rates and minimal overhead. Most of the benefit is achieved by applying a small subset of prioritized recommendations, as shown in Figure~\ref{fig:ablation_curve}, confirming the value of ARRC’s modular, explainable recommendation process and operator-in-the-loop workflow. In fully autonomous mode, ARRC consistently outperforms state-of-the-art autoscalers like VPA, even when enforcing additional constraints.

\subsection{Threats to Validity}
\label{sec:threats}

\textbf{External Validity.}
The generalizability of our results is constrained by the use of a proprietary industrial dataset and production platform, with only relative figures reported for privacy and compliance reasons. This restricts direct reproducibility and limits the ability of others to compare results or replicate findings, a challenge common in industry-facing resource management research. Although ARRC was validated at scale in a multi-region edge--cloud continuum, the evaluation reflects a single-case deployment and may not capture the full diversity of environments, workloads, or organizational practices seen in the broader ecosystem. The absence of open datasets or standardized public benchmarks for complex, multi-tenant edge–cloud resource optimization further complicates cross-study comparison and external validation. Effectiveness may be reduced in contexts with highly unpredictable workloads, insufficient telemetry, or lack of integration with supported platforms such as Kubernetes or OpenStack. Rapid workload changes without sufficient historical data can also challenge the forecasting models and diminish recommendation quality. While autonomous ARRC vs.\ VPA comparisons show recurring efficiency gains, these experiments were limited to one-week windows and single cluster settings; longer-term, multi-cluster, and multi-organization studies are needed for general conclusions and to establish robustness at larger scale.

\textbf{Internal Validity.}
ARRC currently employs established time-series forecasting and heuristic decision logic (see Section ~\ref{sec:approach}), but does not incorporate advanced or adaptive learning models for highly dynamic or bursty workloads. This design choice, motivated by operational simplicity and maintainability, may limit the ability of the system to capture all forms of real-time variability or abrupt pattern shifts. Internal validity may therefore be affected in settings with frequent or unpredictable changes. Additionally, although all results are based on live production data, the absence of an open-source implementation and full dataset limits independent validation of internal processes.

\textbf{Construct Validity.}
The evaluation primarily focuses on CPU and memory utilization, and does not directly quantify broader objectives such as cost-efficiency, performance, or reliability, although ARRC’s recommendation structure is designed to support multi-objective optimization (see Table~\ref{tab:arrc-sample-recs}). While the agent framework is architected to incorporate ISO/IEC 25010 software quality constraints, the current deployment enforces only a subset. Moreover, the study quantifies reduction in operator effort, but does not systematically measure or benchmark operator attitudes, trust, or acceptance; any such observations are informal and anecdotal. The lack of available standards and open benchmarks for explainable, workflow-integrated resource optimization further limits the ability to systematically compare ARRC to other approaches.

\textbf{Summary.}
While ARRC demonstrates strong results in an industrial setting, its broader applicability, adaptability to dynamic workloads, and ability to optimize for multiple, possibly conflicting, objectives merit further study. The results presented should be considered an initial indication, and further validation in larger-scale, multi-environment, and multi-organization scenarios, particularly with public benchmarks and standardized methods, remains an important direction for future work. Continued work on these limitations will help establish ARRC as a general-purpose, robust, and transparent optimization assistant for the edge--cloud continuum, exemplifying maintainable and explainable automation in software engineering practice. %
\section{Discussion}
\label{sec:discussion}

\textbf{Targeted Problem-Solving and Integration.}
Our iterative development and deployment of ARRC highlighted the importance of focusing on clear, well-defined gaps in existing resource management and automation tools. Initial attempts to subsume or replace established systems, such as VPA, HPA, or Cluster Autoscaler \cite{vpa, hpa, clusterautoscaler}, introduced complexity and limited practical value. By instead targeting underserved needs, such as proactive rightsizing for long-lived workloads, multi-objective recommendations, and cross-layer explainability, ARRC achieved broader applicability and easier integration with existing workflows. This practitioner experience suggests that building on proven foundations and addressing specific, actionable gaps is more effective than aiming for wholesale replacement of robust legacy solutions.

\textbf{Simplicity and Workflow Integration.}
While advanced predictive models and complex scheduling algorithms were considered, ARRC's design prioritized transparency, explainability, and integration with familiar operator workflows. Simpler, modular approaches were found to minimize operational risk and improve maintainability, even if they did not always maximize automation or theoretical optimality. Operator effort reduction was quantitatively measured, while other aspects, such as operator trust or acceptance, were only observed informally and not systematically evaluated in this study.

\textbf{Reproducibility and Transparency.}
Despite detailed algorithmic descriptions and pseudocode, practical challenges in data sharing and replication remain a barrier for open research. Privacy, contractual, and organizational constraints prevented release of real operational data, limiting direct external validation. While this paper aims to facilitate broader adoption and re-implementation through technical transparency, the lack of shareable datasets remains an open challenge for the community. Continued advocacy and development of synthetic benchmarking resources will be necessary for advancing reproducibility in this field.

\textbf{Summary.}
In sum, sustainable, operator-centric optimization in the edge–cloud continuum benefits from targeted problem-solving, transparent and auditable automation, and measurable operator impact. While some practitioner insights, such as operator attitudes and preferences, are only supported by informal observations, the quantitative results on efficiency and operator effort are empirically grounded. These findings informed the development and deployment of ARRC and may provide useful guidance for others pursuing robust, maintainable, and transparent automation in production-scale environments. %
\section{Conclusion and Future Work}\label{sec:conclusion}

This work addresses the persistent challenge of sustainable, maintainable resource optimization across the edge--cloud continuum by introducing the Automatic Recommender for Resource Configurations (ARRC), a modular recommender architecture grounded in software engineering and reliability design principles. ARRC unifies proactive, cross-layer optimization for virtual machines, containers, and security domains, integrating directly with existing platforms such as Kubernetes and OpenStack. The system continuously analyzes workload telemetry, applies forecasting pipelines, and generates actionable, explainable recommendations for resource right-sizing, security, and manageability. Recommendations are delivered through established operator workflows, including tickets and GitOps pull requests, facilitating transparent and incremental adoption.

Empirical results from a multi-month, multi-region industrial deployment demonstrate that ARRC substantially improves compute utilization, achieving up to $7.7\times$ gains over baseline, with most benefits realized through a small fraction of prioritized recommendations. ARRC reduces operator workload by more than $50\%$, cuts average handling time per optimization from $27$ to $7$ minutes, and decreases total monthly operator effort from $19$ to $8.4$ hours, while maintaining low error rates and minimal operational overhead.

ARRC’s key contribution is an extensible architecture that prioritizes modularity, explainability, and maintainability, enabling transparency and the ability to inspect and understand recommendations. All core algorithms, agent interfaces, and evaluation methods are documented for reproducibility and future extension by the community.

Looking ahead, we plan to extend ARRC’s applicability to more dynamic workloads and platforms, broaden optimization objectives (e.g., cost, sustainability), integrate adaptive learning agents for improved responsiveness, and enhance reproducibility through public benchmarks and replication studies. While the ARRC codebase is not open source, this paper provides the technical detail needed for re-implementation and scientific review.

In summary, ARRC demonstrates that structured, explainable, and workflow-integrated recommendation architectures can deliver sustainable efficiency and maintainability improvements in production-scale edge--cloud continuum systems. We hope this work inspires continued progress in transparent, operator-centric resource management and reinforces the value of principled, maintainable automation in real-world software engineering practice. 

\section*{Availability}

The Automatic Recommender for Resource Configurations (ARRC) is accessible for
both research and commercial purposes.  ARRC is distributed under license, for
further information please contact the authors or their employer.

\bibliographystyle{plain}
\bibliography{literature}

\end{document}